\documentclass[11pt]{article}
\pdfoutput=1
\usepackage{amsmath, amsfonts, amssymb}
\usepackage{comment}
\usepackage{graphicx}
\usepackage{psfrag}
\usepackage{amsthm}
\usepackage[usenames,dvipsnames,svgnames,table]{xcolor}
\usepackage{enumerate}
\usepackage{tcolorbox}
\usepackage{mathtools}
\usepackage{soul}
\usepackage{subfigure}
\usepackage{mathrsfs}  
\usepackage{a4wide}
\usepackage{booktabs}
\usepackage{tikz}
\usepackage{tikz-cd}
\usepackage{makecell}
\usepackage{color}
\definecolor{dark-gray}{gray}{0.20}
\definecolor{gray}{gray}{0.30}
\definecolor{light-gray}{gray}{0.80}
\definecolor{dark-red}{rgb}{0.7,0,0}
\definecolor{dark-green}{rgb}{0.1,0.4,0}
\definecolor{dark-blue}{rgb}{0.3,0.3,0.7}
\definecolor{light-blue}{rgb}{0.8,0.8,1}
\definecolor{swamp}{RGB}{240, 199, 197}

\usepackage{pifont}
\usepackage{setspace}

\newcommand{\be}{\begin{equation}}
\newcommand{\ee}{\end{equation}}

\def\be{\begin{equation}}
\def\ee{\end{equation}}
\def\bea{\begin{eqnarray}}
\def\eea{\end{eqnarray}}

\newcommand{\dd}{\mathrm{d}}
\def\simleq{\; \raise0.3ex\hbox{$<$\kern-0.75em
		\raise-1.1ex\hbox{$\sim$}}\; }
\def\simgeq{\; \raise0.3ex\hbox{$>$\kern-0.75em
		\raise-1.1ex\hbox{$\sim$}}\; }

\numberwithin{equation}{section}

\usepackage{jheppub} %jheppub already loads hyperref; load it again and it causes trouble
\hypersetup{
	colorlinks=true,
	linkcolor=dark-blue,
	citecolor=dark-red,
	urlcolor=dark-green,
	linktoc=page,
	pageanchor=false
}

\title{\centering
No Asymptotic Acceleration\\
without Higher-Dimensional de Sitter Vacua\\
}

\author{Arthur Hebecker$^1$,}
\author{Simon Schreyer$^1$,}
\author{and Gerben Venken$^1$}

\affiliation{$^1$ Institute for Theoretical Physics, Heidelberg University,\\
	Philosophenweg 19, 69120 Heidelberg, Germany} 
\emailAdd{a.hebecker@thphys.uni-heidelberg.de}
\emailAdd{s.schreyer@thphys.uni-heidelberg.de}
\emailAdd{g.venken@thphys.uni-heidelberg.de}

\abstract{
There has recently been considerable interest in the question whether and under which conditions accelerated cosmological expansion can arise in the asymptotic regions of field space of a $d$-dimensional EFT. We conjecture that such acceleration is impossible unless there exist metastable de Sitter vacua in more than $d$ dimensions. That is, we conjecture that `Asymptotic Acceleration Implies de Sitter' (AA$\Rightarrow$DS). Phrased negatively, we argue that the $d$-dimensional `No Asymptotic Acceleration' conjecture (a.k.a.~the `strong asymptotic dS conjecture') follows from the de Sitter conjecture in more than $d$ dimensions. The key idea is that the relevant field-space asymptotics almost always correspond to decompactification and that the only positive energy contribution which decays sufficiently slowly in this regime is the vacuum energy of a higher-dimensional metastable vacuum. This result is in agreement with recent Swampland bounds on the potential in the asymptotics in field space from e.g.~the species bound, but is significantly more constraining. We further note that for our asymptotic decompactification limits based on higher-dimensional de Sitter, the Kaluza-Klein scale always falls below the Hubble scale asymptotically.
In fact, this occurs whenever $|V'|/V \leq 2 \sqrt{(d+k-2)/k(d-2)}$ asymptotically, with $k$ the number of decompactifying internal directions. This is steeper than what is needed for accelerated expansion.}

\newtheorem{conjecture}{Conjecture}
\begin{document}
%%%%%%%%%%%%%%%%%%%%%%%%%%%%%
%%%%%%%%%%%%%%%%%%%%%%%%%%%%%

\makeatletter
\let\old@fpheader\@fpheader

\makeatother

\maketitle

%\tableofcontents

%%%%%%%%%%%%%%%%%%%%%%%%%%%%%
%%%%%%%%%%%%%%%%%%%%%%%%%%%%%
\section{Introduction}\label{intro}
%%%%%%%%%%%%%%%%%%%%%%%%%%%%%
%%%%%%%%%%%%%%%%%%%%%%%%%%%%%
In recent years, doubt has been cast on whether it is possible to construct de Sitter vacua in string theory \cite{Danielsson:2018ztv,Obied:2018sgi}. A related, maybe simpler question is whether one can exclude accelerated cosmological expansion in the asymptotic regime of field space, e.g. by appealing to entropy or species scale arguments \cite{Ooguri:2018wrx, Hebecker:2018vxz}. The possibility of asymptotic acceleration has continued to enjoy significant attention \cite{Grimm:2019ixq,Calderon-Infante:2022nxb, vandeHeisteeg:2023uxj, Shiu:2023nph,Shiu:2023rxt,Cremonini:2023suw}. 

In this note we present an argument that achieving accelerated expansion asymptotically in the field space of a $d$-dimensional theory is precisely as hard as constructing a higher-dimensional de Sitter vacuum. It then seems that the direct search for $d$-dimensional de Sitter vacua in the interior of moduli space is the simpler and more straightforward option to establish accelerated expansion in string theory.
We note that our investigation is motivated largely at the conceptual level and is hence distinct from the also very important but difficult quest for realistic quintessence models in string theory, see e.g.~\cite{Cicoli:2011yy, Cicoli:2012tz, Cicoli:2018kdo, Olguin-Trejo:2018zun, Hebecker:2019csg}.

Let us pause to recall why realizing asymptotic expansion in string theory is important independently of quintessence-type phenomenological applications: It is conceivable and maybe expected that any fundamental difficulties of constructing de Sitter space or cosmological inflation in string theory are related to the presence of cosmic horizons.
At the conceptual level one then expects that it should be equally hard to achieve de Sitter space or accelerated expansion in a theory of quantum gravity. This is of course not a new point, see e.g. \cite{Fischler:2001yj,Hellerman:2001yi}. Crucially, if one could realize asymptotic accelerated expansion in string theory, it would become clear that cosmic horizons are {\it not} forbidden by quantum gravity as a matter of principle. Then one could be more optimistic about stringy de Sitter models, even if those are not fully controlled.
However, our analysis appears to suggests that the strategy outlined above is problematic.

Before stating our main result, let us first define more precisely what we mean by asymptotic acceleration and how this is related to the presence of cosmic horizons. For this, we first note that acceleration (defined by $\ddot{a}>0$, where $a$ is the scale factor) is equivalent to the equation-of-state parameter fulfilling $w<w_\text{crit}$, where $w_\text{crit}=-(d-3)/(d-1)$. Next, a cosmic horizon exists if the integral 
    \begin{equation}
        d_h \sim \int_{t_i}^{t_f} \frac{\dd t}{a(t)}
        \label{horizon}
    \end{equation}
stays finite at $t_f\to \infty$. It immediately follows that cosmic horizons exist if $w(t)\to w_\text{crit}-\epsilon$, for positive constant $\epsilon$. Indeed, in this case the large-$t$ behavior of the scale factor is $a\sim t^n$ with $n=1/[1-\epsilon(d-1)/2]$, such that our integral converges. If we now define asymptotically accelerating cosmologies by the requirement that $w(t)< w_\text{crit}-\epsilon$ at $t\to\infty$ for some positive $\epsilon$, then it is obvious that asymptotic acceleration implies horizons. Ruling out asymptotic acceleration as defined above rules out a large class of cosmologies with horizons\footnote{Note that our definition of asymptotic acceleration does not exclude models where $w(t)\to w_\text{crit}$ from below, such that $\ddot{a}>0$ asymptotically but the approach of the critical value is so slow that the \eqref{horizon} converges. Thus, there is room for softening our definition of asymptotic acceleration to capture these cases. We leave that to future work. 
Furthermore, as we will comment on in more detail in Sect.~\ref{sec:externaleffects}, the models of cosmic acceleration that rely on negative spatial curvature ($k=-1$) are not counted as asymptotically accelerating by our definition since they do not lead to horizons as recently discussed in \cite{Andriot:2023wvg}.}

Given the above definition, we may now summarize our main results in the following central conjecture:

\begin{conjecture}[`Asymptotic Acceleration Implies de Sitter' or `AA$\Rightarrow$DS']\label{mainconjecture}
Consider a $d$-dimensional (effective) theory of quantum gravity ($d>2$).
If this theory realizes accelerated expansion through rolling scalars at asymptotically large distance in field space, then this theory is based on the compactification of a $(d+k)$-dimensional theory with positive vacuum energy. Given that no quantum gravity theories exhibit dS vacua in their fundamental dimension, the underlying $(d+k)$-dimensional de Sitter model is itself the result of a compactification. Further, whenever the potential asymptotically accelerates, this implies that the Kaluza-Klein scale becomes lighter than the Hubble scale asymptotically.
\end{conjecture}

Actually, a stronger result holds: The Kaluza-Klein scale becomes lighter than the Hubble scale asymptotically whenever $|V'|/V \leq 2 \sqrt{(d+k-2)/k(d-2)}$ and $V>0$. This bound requires the potential to be steeper than what is needed for acceleration in order to achieve scale-separation. In such a situation there is no longer a cosmic horizon, but the Hubble scale still sets the curvature scale of the external dimensions.

Our conjecture deals with asymptotic accelerated expansion. Achieving a finite period of accelerated expansion in the interior of moduli space already runs into all the control issues of obtaining de Sitter vacua. A central point of our conjecture is that one cannot achieve accelerated expansion in some clean parametrically controlled way by considering the asymptotics of moduli space. Those asymptotics which give accelerated expansion rely on higher-dimensional de Sitter vacua and so one must deal with the control issues of de Sitter model building.

In the Swampland program, there exists a whole miasma of conjectures about positive potentials in string theory \cite{Danielsson:2018ztv,Obied:2018sgi,Ooguri:2018wrx,Garg:2018reu, Denef:2018etk, Andriot:2018mav,Hebecker:2018vxz,Junghans:2018gdb,Hebecker:2019csg,Cicoli:2018kdo,Bedroya:2019snp,Rudelius:2021oaz,Rudelius:2021azq,Andriot:2020lea,Cicoli:2021skd,Cicoli:2021fsd}. 
Most closely related to our conjecture is the `No Asymptotic Acceleration' conjecture (a.k.a.~the `strong asymptotic dS conjecture') \cite{Rudelius:2021oaz,Rudelius:2021azq}, claiming that the scalar potential should obey $|\nabla V|/V \geq 2 / \sqrt{d-2}$ in $d$-dimensional Planck units. But even if restricted to the asymptotics of moduli space (which is almost certainly necessary for the reasons explained in \cite{Hebecker:2018vxz}), this conjecture is stronger than ours since it forbids accelerated expansion. Our weaker conjecture does a priori not forbid anything -- it states that to achieve asymptotic accelerated expansion, one must be able to construct de Sitter vacua in higher dimensions. However, it turns out that in the resulting cosmologies the KK scale eventually falls below the cosmological curvature scale. One may then argue that accelerated expansion is forbidden after all, at least if one insists that a $d$-dimensional EFT description is only useful below the KK scale.

To argue for our conjecture, we start in Sec.~\ref{sec:branecalc} by considering a $d$-dimensional theory in the asymptotic regime where $k$ further dimensions decompactify. We argue that, in this situation, accelerated expansion can only be achieved if the underlying $(d+k)$-dimensional theory has a positive cosmological constant. Referring to the Emergent String Conjecture \cite{Lee:2018urn, Lee:2019wij, Cota:2022yjw,Wiesner:2022qys}, we explain why the above analysis of the decompactification limit suffices to reach our conclusions in {\it any} asymptotic limit. Independent arguments for this as well as further reasoning supporting our conjecture are provided in Sec.~\ref{sec:limits}. 
Finally, Sec.~\ref{sec:discussion} discusses implications of our conjecture in the context of string model-building and the swampland.
This note has two appendices. In App.~\ref{sec:expansioncond} we discuss recent results of \cite{Calderon-Infante:2022nxb,Shiu:2023nph,Shiu:2023rxt} on asymptotic acceleration with multiple fields. In particular we present a theorem of \cite{Shiu:2023nph,Shiu:2023rxt} in a form which we will rely on in the main text. These results provide a sufficient bound on certain scalar potentials to forbid accelerated expansion. App.~\ref{sec:appendix} presents technical calculations on the canonical normalization of the volume modulus and the scaling of potential terms. Finally, App.~\ref{sec:axionapp} comments on scale separation in the context of the acceleration mechanism of \cite{Cicoli:2020cfj,Cicoli:2020noz,Brinkmann:2022oxy}, also discussed in Sec.~\ref{sec:limits}.

The reader may immediately object to our conjecture by noting that \cite{Calderon-Infante:2022nxb} appears to provide an explicit counterexample. However, as discussed in Sec.~\ref{sec:limits} (cf.~also the very recent paper \cite{Cremonini:2023suw}), upon closer inspection this is not actually a counterexample.

%%%%%%%%%%%%%%%%%%%%%%%%%%%%%
%%%%%%%%%%%%%%%%%%%%%%%%%%%%%
\section{Scaling of positive potential terms in the decompactification limit}
\label{sec:branecalc}
%%%%%%%%%%%%%%%%%%%%%%%%%%%%%
%%%%%%%%%%%%%%%%%%%%%%%%%%%%%

%%%%%%%%%%%%%%%%%%%%%%%%%%%%%
%%%%%%%%%%%%%%%%%%%%%%%%%%%%%
\subsection{The decompactification limit and acceleration}
%%%%%%%%%%%%%%%%%%%%%%%%%%%%%
%%%%%%%%%%%%%%%%%%%%%%%%%%%%%

In order to see whether asymptotic accelerated expansion is possible, we wish to know what the flattest positive potential is that can be achieved asymptotically in field space in a string compactification. The Emergent String Conjecture, as developed in \cite{Lee:2018urn, Lee:2019wij, Cota:2022yjw,Wiesner:2022qys}, implies that any asymptotic limit in string theory in less than ten dimensions is a limit in which a tower of KK modes becomes light (cf.~Proposal 1\footnote{According to this proposal, any infinite distance limit is either a decompactification limit or a limit in which a string becomes light. In the case where a string becomes light, a tower of KK modes always also becomes light, with its mass scale parametrically the same as that of the string tower.} and the preceding discussion in \cite{Lee:2019wij}). We can then interpret any asymptotic limit in a string compactification as a decompactification limit. This is not necessarily a decompactification to 10D.\footnote{Throughout this section, we find it convenient to assume that the fundamental theory before compactification is the 10D critical string. However, both our analysis and our conclusions remain unchanged if we start from 11D M-theory or any other fundamental theory of quantum gravity.} 
In other words, part of the internal dimensions may remain compactified or at least parametrically smaller than the largest radius. 
Note that the only property of the decompactification limit relevant to our analysis is that at least one of the (potentially several, different) compactification scales, measured in $d$ dimensional Planck units, rolls to infinity during late cosmology. While this goes together with the lowest KK scale falling below any fixed cutoff, this aspect is not important to us at the present point. All we need is the fact that one radius $R$ approaches infinity at late times, $R\to\infty$. 
In the rest of this section, we will take it for granted that, to study the possibility of asymptotic accelerated expansion, it suffices to study decompactification limits. In Sec.~\ref{sec:limits} we will provide our own motivation for this assumption and deal with potential counterexamples and caveats.

Let us now place bounds on the dominant term in the scalar potential. If this term is $\sim\exp(-\gamma\phi)$, with $\phi$ a canonically normalized scalar, a sufficient condition to forbid asymptotic accelerated expansion in $d>2$ (to which we restrict ourselves) is
\begin{equation}
\label{eq:acccondmain}
    \gamma \geq \frac{2}{\sqrt{d-2}} \equiv \gamma_{\text{acc}}\,,
\end{equation}
where $d$ is the number of non-compact dimensions. This can be derived straightforwardly (see e.g. \cite{Calderon-Infante:2022nxb,Shiu:2023nph,Shiu:2023rxt}).

\begin{table}[!h]\centering
	\caption{Most important sources for potential energy after compactification from $d+k$ to $d$ dimensions. The second line displays the scaling with the common radius $R$ of the compact directions.}
	\vspace{.3cm}
	\label{tab:scalingsmain}
		\begin{tabular}{c|cccc}
		\toprule
		    candidate & $p$-branes & $l$-form flux & curvature & loops\\
			\hline
            \midrule
			scaling in $d$ dim Brans-Dicke frame& $R^{p+1-d}$& $R^{k-2l}$&$R^{k-2}$ &$R^{-d}$\\
			\bottomrule
	\end{tabular}
\end{table}

We consider a compactification from $d+k$ to $d$ dimensions, assuming that asymptotically all $k$ internal dimensions have a radius $\sim R$, with $R$ rolling to infinity. 
Note that in this section we neglect the existence of a rolling potential of the form $V\sim \exp(-\gamma\varphi)$ in the $(d+k)$-dimensional theory where $\varphi$ is a canonically normalized scalar. This clearly provides a loophole to our conjecture if $\gamma>0$ is allowed to be arbitrarily small. However, as we prove in Sec.~\ref{sec:limits} there is a lower bound on $\gamma$ in string theory deriving from the fact that in 10D/11D the slope of such a rolling potential is bounded. This lower bound on $\gamma$ does not allow for asymptotic accelerated expansion in $d$ dimensions induced by a rolling potential in the $(d+k)$-dimensional theory.

The key sources which can contribute to the resulting scalar potential in the $d$-dimensional theory are listed in Table~\ref{tab:scalingsmain}. These are spacetime-filling $p$-branes wrapped on appropriate cycles of the compact space, the kinetic term of an $l$-form flux, the leading curvature term (Einstein-Hilbert term), and the loop effect or Casimir energy of massless fields. Note that here the term $p$-brane is not restricted to the fundamental branes of the known superstring theories but includes any $(p+1)$-dimensional defect of the $(d+k)$-dimensional (effective) theory. The scaling of the resulting potential terms with $R$
in $d$-dimensional Brans-Dicke frame is also displayed in the Table. We should emphasize that our calculations here are not at all new as similar analyses have appeared e.g.~in \cite{Hertzberg:2007wc,Rudelius:2021oaz,Rudelius:2021azq, Etheredge:2022opl, Rudelius:2022gbz}. We only differ in the interpretation of the results.
We focus on the $R$-dependence of the potential terms and ignore the possible dependence on further parameters of the $(d+k)$-dimensional theory. Such parameters include
$g_s$, shape moduli, or internal radii of the compactification leading to the $(k+d)$-dimensional theory. If they are not stabilized then, as we will argue in 
Sec.~\ref{sec:limits}, they only provide additional directions to roll and lower the potential energy. They will hence not save situations in which the $R$-direction is too steep.

It is clear from Table \ref{tab:scalingsmain} that one achieves the flattest $d$-dimensional scalar potential ($\sim R^k$) either from a $p$-brane filling all $d+k$ dimensions or $0$-form flux. Both of these cases correspond to a cosmological constant in the $(d+k)$-dimensional theory. Going to Einstein frame and rewriting $R$ in terms of the corresponding canonically normalized scalar $\phi$ (see App.~\ref{sec:appendix}) provides a $d$-dimensional scalar potential of the form
\begin{equation}\label{pothigherdimcc}
    V\sim \exp{\left(-\frac{2}{\sqrt{d-2}}\sqrt{\frac{k}{k+d-2}}\, \phi\right)}\,.
\end{equation}
By comparison with \eqref{eq:acccondmain} one sees that such a potential always yields asymptotic accelerated expansion.

The next-flattest potential ($\sim R^{k-1}$) stems from a codimension-1 brane in the $(d+k)$-dimensional theory, i.e.~$p=d+k-2$. In Einstein frame this yields a potential
\begin{equation}\label{potcodimone}
    V\sim \exp{\left(-\frac{2}{\sqrt{d-2}}\, \frac{k+d/2-1}{\sqrt{k(k+d-2)}}\, \phi\right)}\,.
\end{equation}
One sees from \eqref{eq:acccondmain} that this will never give asymptotic accelerated expansion.

This is one of our central results, justifying Conjecture \ref{mainconjecture}. Section~\ref{sec:limits} and App. \ref{sec:expansioncond} provide further justification for the assumptions which went into obtaining this conclusion.

Leaving details to this Appendix,
let us state already now why multifield rolling does not falsify our conjecture: The key reason is the universality of the volume modulus $R$. By this we mean that \textit{every} term of the resulting effective potential acquires an $R$ dependence. In part, this is due to the $R$-dependence of the 4d Planck mass, in part it comes about because each of the underlying $(d+k)$-dimensional effects (e.g.~flux or Casimir energy) dilutes with growing $R$. If every term is steeply falling and the conditions outlined in App.~\ref{sec:expansioncond} apply, then asymptotically the resulting dependence of the full potential can not be less steep. This statement about the overall steepness in one distinguished field direction does \textit{not} depend on the presence of further fields.

We finally note that \eqref{pothigherdimcc} provides the flattest possible potential when $k=1$, in which case the potential exactly saturates the Trans-Planckian Censorship bound \cite{Bedroya:2019snp} which asymptotically in moduli space states that the exponent should obey $\gamma \geq 2/\sqrt{(d-1)(d-2)}$. A corollary of our results is then that for string compactifications the TCC holds in the asymptotics of moduli space.

%%%%%%%%%%%%%%%%%%%%%%%%%%%%%
%%%%%%%%%%%%%%%%%%%%%%%%%%%%%
\subsection{Kaluza-Klein scale becoming lighter than Hubble scale}\label{sec:EFTbreakdown}
%%%%%%%%%%%%%%%%%%%%%%%%%%%%%
%%%%%%%%%%%%%%%%%%%%%%%%%%%%%

It is natural to ask whether in the decompactification limit the Kaluza-Klein scale asymptotically becomes lighter than the Hubble scale. Note that the Kaluza-Klein scale becoming lighter than the Hubble scale does not imply a complete breakdown of the $d$-dimensional effective theory, for that the species scale would have to become lighter than the Hubble scale. However, once $H\geq1/R$, a naive $d$-dimensional picture is no longer the correct one to treat certain cosmic properties such as cosmic horizons. As an example, the geometry of a four-dimensional Nariai black hole is $dS_2 \times S^2$, where the $dS_2$ and $S^2$ are both of the same length scale. From a naive $dS_2$ perspective the horizon is pointlike and the entropy is an order one number. The correct four-dimension perspective sees the $S^2$ at the horizon and correctly assigns an entropy proportional to the horizon area in four dimensions.

Let us suppose both scalar kinetic energy and potential energy contribute to cosmic evolution without assumption about which dominates. In this case\footnote{Note that this equation assumes scalars with canonical kinetic terms and the absence of curvature in the external directions. We will return to these assumptions in Sec. \ref{sec:externaleffects}.}
\begin{equation}
\label{eq:hubblefromenergy}
    H \sim \frac{1}{M_{p,d}^{\frac{d-2}{2}}} \sqrt{M_{p,d}^{d-2}\Dot{\phi}^2 + V} \geq  M_{p,d} \sqrt{\frac{V}{M_{p,d}^{d}}}\,.
\end{equation}
Setting $V/M_{p,d}^{d}\sim \exp(-\gamma \phi)$ and using eq. \eqref{eq:planckconversion}, one may write the ratio of Hubble and KK scale as
\begin{equation}
    \frac{H}{1/R} \gtrsim \frac{M_{p,d}}{1/R}\exp(-\gamma \phi /2) \sim M_{p,d+k}^{\frac{d+k-2}{d-2}} R^{1+\frac{k}{d-2}} \exp(-\gamma \phi /2)\,.
\end{equation}
Expressing $R$ in terms of the canonical scalar via eq. \eqref{eq:volumecanon} one finds
\begin{equation}
    \frac{H}{1/R} \,\,\gtrsim\,\,  \exp\left[\left(\frac{-\gamma}{2}+\sqrt{\frac{d+k-2}{k(d-2)}}\,\right)\phi\,\right]\,.
\end{equation}
Hence, to avoid the cosmic scale exceeding the KK scale asymptotically, one must have $\gamma/2 > \sqrt{(d+k-2)/k(d-2)}$. Combining this with the relation $\gamma_{\rm acc}=2/\sqrt{d-2}$ gives
\begin{equation}
    \gamma > 2 \sqrt{\frac{d+k-2}{k(d-2)}} > \gamma_{\text{acc}}\,.
\end{equation}
These inequalities imply that the for asymptotically accelerating cosmologies, and in fact even for somewhat steeper potentials, one asymptotically always has $H \geq 1/R$

%%%%%%%%%%%%%%%%%%%%%%%%%%%%%
%%%%%%%%%%%%%%%%%%%%%%%%%%%%%
\section{Covering additional limits and contributions to acceleration}
\label{sec:limits}
%%%%%%%%%%%%%%%%%%%%%%%%%%%%%
%%%%%%%%%%%%%%%%%%%%%%%%%%%%%

In the present section, we want to tie up several loose ends left in Sec.~\ref{sec:branecalc}. 

%%%%%%%%%%%%%%%%%%%%%%%%%%%%%
\subsection{The not-so-many asymptotic limits of string compactifications}
%%%%%%%%%%%%%%%%%%%%%%%%%%%%%

First, we recall that we had simply referred to the Emergent String Conjecture literature \cite{Lee:2018urn,Lee:2019wij,Cota:2022yjw,Wiesner:2022qys} to claim that every infinite-distance limit with $d<10$ can be thought of as a decompactification limit. We now want to explicitly argue for this point. Second, we want to prove that allowing for an exponentially falling potential in the $(d+k)$-dimensional theory does not endanger our conjecture. Third, we comment on the apparent counterexample of \cite{Calderon-Infante:2022nxb}. Arguments that multiple rolling moduli do not affect our conjecture are discussed in App.~\ref{sec:expansioncond}.

Consider first the case where only the dilaton is rolling. As shown in~\cite{Shiu:2023nph, Shiu:2023rxt} using standard lagrangians, no asymptotic acceleration arises in this case. We note in addition that if the radius $R$ is fixed in Planck units, then $g_s\to 0$ can be reformulated as $R/\sqrt{\alpha'}\to 0$. If the compact geometry involves $S^1$s such that T-duality can be applied, one may think in terms of the diverging $T$-dual radius, consistently with our decompactification assumption. We expect that, for geometries where no T-dual formulation is possible, $R/\sqrt{\alpha'}\to 0$ either does not correspond to an infinite distance limit (see also below), or that within a finite distance an effective Einstein gravity description breaks down even at the energy scale set by the curvature of the external dimensions, see e.g.~\cite{Gaberdiel:2014cha}.

Having thus excluded situations where {\it only} the dilaton rolls, we can now dismiss the dilaton altogether. Indeed, if $g_s\to 0$ with $R/\sqrt{\alpha'}$ staying finite, the radius diverges in terms of 4d Planck units and we are thrown back to our decompactification limit. The additional presence of a rolling dilaton should be harmless according to our arguments in App.~\ref{sec:expansioncond}.

We now turn to geometric moduli, distinguishing between the three cases where, at infinite distance, the compact volume goes to infinity, zero or to a finite value.

The infinite-volume limit is, of course, precisely the decompactification limit which we covered in Sec.~\ref{sec:branecalc}. 

The zero-volume limit can be mapped to a decompactification limit by applying one or more T-dualities. This is the same argument used above for $g_s\to 0$. It may also work for geometries possessing no explicit torus factors but instead appropriate fibrations (cf. Calabi-Yau mirror symmetry). 
If the compact volume shrinks in a way which does not allow for a T-dual infinite-volume interpretation, we expect that the distance in moduli space leading to this point is finite. Examples are the singular limit of the deformed or blown-up conifold or the shrinking of the $S^3$ base of a Calabi-Yau, viewed as a $T^3$ fibration.

Finally, we expect that limits in which the compact volume stays finite while a geometric modulus rolls to infinity are always decompactification limits. One obvious example are Calabi-Yaus which are appropriately fibred, such that e.g. some Kahler modulus governing the base goes to infinity while at the same time the fibre shrinks. Clearly, in the infinite distance limit the base decompactifies. Another example is that of the large-complex-structure limit, where the $T^3$ fibre of the Calabi-Yau shrinks and, once again its base, this time an $S^3$, decompactifies. In fact, in both cases eventually the small fibre has to be T-dualized and decompactifies as well. We expect the above logic to also apply to a `partial large-complex-structure limit' where one of the moduli $z^i$ diverges. The reason is that the volume of the mirror Calabi-Yau geometry involves this $z^i$ and will hence diverge.

This logic then also applies to the apparent counterexample given in \cite{Calderon-Infante:2022nxb} where the dilaton and a complex structure modulus are rolling which seemingly allows for accelerated expansion. However, in that example Kahler moduli are entirely neglected. Including Kahler moduli we expect that in general the volume modulus also rolls preventing accelerated expansion. If all Kahler moduli could be stabilized in such a way that asymptotically $V>0$, we expect that the volume of the mirror Calabi-Yau decompactifies. This agrees with the finding of \cite{Calderon-Infante:2022nxb} that after mirror symmetrizing to the type IIA side no counterexample to the No Asymptotic Acceleration conjecture \cite{Rudelius:2021oaz,Rudelius:2021azq} could be found.

This still leaves one final option: One could attempt to realize an asymptotically accelerating set-up along the lines of \cite{Calderon-Infante:2022nxb} in a geometry with no Kahler moduli, i.e.~$h^{1,1}=0$. In this case one obviously need not worry how Kahler moduli affect the potential. However, such a setup should be mirror dual to a setup with only Kahler moduli. In this case one again recovers a situation which should be equivalent to a decompactification limit. We expect that our conjecture then excludes acceleration and hence, indirectly, the possibility of realizing the proposal of \cite{Calderon-Infante:2022nxb} at $h^{1,1}=0$. This expectation agrees with the recent results of \cite{Cremonini:2023suw} where the authors tried to mimic the counterexample of \cite{Calderon-Infante:2022nxb} in full-fledged stringy models with $h^{1,1}=0$ without finding trajectories in field space that allow for asymptotic accelerated expansion. 

Independent of the preceding points, if the proposal of \cite{Calderon-Infante:2022nxb} does accelerate, then by considering it in a duality frame where the geometry decompactifies, it follows from Sec.~\ref{sec:EFTbreakdown} asymptotically one will have $H \geq 1/R$.

\subsection{Higher-dimensional potential}

Let us now allow for a potential of the form $V\sim\exp(-\gamma\varphi)$ in the $(d+k)$-dimensional theory which was disregarded in Sec.~\ref{sec:branecalc}. If 
\begin{equation}
     \gamma \geq \gamma_{\text{acc},d+k} = \frac{2}{\sqrt{d+k-2}}\,,
\end{equation}
the potential does not allow for asymptotic acceleration in $d$ dimensions, as observed in~\cite{Rudelius:2021oaz}. By contrast, if $\gamma < \gamma_{\text{acc},d+k}$, then we already have accelerated expansion in the $(d+k)$-dimensional theory. 
In other words, compactifying a theory with exponential potential to $d$ dimensions produces asymptotic acceleration only if the underlying $(d+k)$-dimensional theory already had asymptotic acceleration. This argument can be applied iteratively until we reach the fundamental dimension of the theory, i.e. $d=10$ or 11 string theory. 
In this case there is either no potential (type IIB, M-theory, heterotic) or the potential does not allow for acceleration. The latter occurs in massive type IIA, the non-SUSY $O(16)\times O(16)$ heterotic string \cite{Alvarez-Gaume:1986ghj}, or the supercritical string with a linear dilaton \cite{Hellerman:2006nx}. We hence conclude our that conjecture is not endangered by adding exponential potentials to the allowed energy sources.

%%%%%%%%%%%%%%%%%%%%%%%%%%%%%
\subsection{External energy sources}
\label{sec:externaleffects}
%%%%%%%%%%%%%%%%%%%%%%%%%%%%%
Until now we have considered contributions to the $d$-dimensional equations of motion coming from sources in the compactified directions which one can effectively treat as part of the $d$-dimensional scalar potential. One can also consider sources which one can {\it not} treat as part of the $d$-dimensional potential. We will focus on two such sources and show that they do not lead to  accelerated expansion in the strict asymptotic limit. In both cases it is however possible to achieve accelerated expansion for an arbitrarily large finite period of time, asymptoting to an equation of state $w=-(d-3)/(d-1)$. The latter is the borderline case between accelerating and non-accelerating cosmologies. Crucially, in a set-up with such asymptotics all cosmic horizons eventually recede and all observers come into causal contact with each other.

First, one may consider curvature in the external dimensions. For homogeneous negative curvature, one has an equation of state $w=-(d-3)/(d-1)$. As just noted, this does not provide acceleration but rather represents the borderline between accelerating and non-accelerating expansion. In \cite{Wohlfarth:2003kw,Russo:2018akp,Marconnet:2022fmx}, string compactifications with negative curvature in the external dimensions were shown to realize cosmologies with accelerated expansion at arbitrarily large finite time. Such a behaviour is not unexpected and can be understood as a deviation from the borderline case, realized by the negative curvature. The origin of such a deviation may be additional sources, for example a potential which is too steep to accelerate by itself. Even if acceleration is present at any finite time, such cosmologies always asymptote to $w=-(d-3)/(d-1)$. In the strict asymptotic limit such cosmologies then do not have cosmic horizons and do not accelerated.

Second, the extremely interesting idea of employing axions in the $d$-dimensional EFT to obtain acceleration was suggested and explored in \cite{Cicoli:2020cfj,Cicoli:2020noz,Brinkmann:2022oxy}. String-theoretically, axions arise from $q$-form gauge fields in the $(d+k)$-dimensional theory, which upon integration over a $q$-cycle in the $k$ internal dimensions realize a scalar field in the non-compact spacetime. Kinetic energy from these axions can contribute to cosmic evolution. In $d$-dimensional Einstein frame, the kinetic terms of the axion $c$ and its dual $(d-1)$-form ${\cal H}_{d-1}$ scale as
\begin{equation}
\label{eq:axionkin}
    \exp\left(-2 k_1 \phi \right) \left(\partial c\right)^2 \quad \xrightarrow{\text{dualizing}}\quad \exp\left(2k_1 \phi \right) \left( {\cal H}_{d-1}\right)^2\,, \quad k_1 = \sqrt{\frac{k(d-2)}{k+d-2}}\frac{q}{k}\,.
\end{equation}
As before, $\phi$ denotes the canonically normalized volume modulus. For an FLRW cosmology with scale factor $a$, one has that ${\cal H}_{d-1} \sim 1/a^{d-1}$ as it homogeneously fills all spatial directions. The field-strength energy for a $(d-1)$-form then scales as $\exp(2k_1\phi)/a^{2(d-1)}$. This affects the rolling of $\phi$ like an exponentially growing potential, albeit one the prefactor of which decays with increasing $a$. 

The resulting cosmology is now determined by the interplay between the potential for $\phi$, driving it to infinity, and the $(d-1)$-form field strength term, slowing it down as explained above. The two terms must be decaying at the same rate, $V\sim \exp(2k_1\phi)/a^{2(d-1)}$. We now assume an explicit potential $V \sim \exp(-k_2 \phi)$ and demand that the resulting cosmology is borderline-non-accelerating, i.e.~that the energy density scales as $\sim 1/a^2$.
This corresponds to an equation of state parameter $w=-(d-3)/(d-1)$. We obtain
\begin{equation}
e^{-k_2\phi} \sim e^{2k_1\phi}/a^{2(d-1)}\sim 1/a^2\,.
\end{equation}
It immediately follows that $\exp(k_2\phi)\sim a^2$, $\exp(k_1 \phi)\sim a^{d-2}$, and hence $k_1/k_2=(d-2)/2$. To get acceleration, the kinetic-term prefactor must be steeper or the potential flatter, i.e. 
$k_1/k_2>(d-2)/2$. This agrees with the analysis of \cite{Cicoli:2020cfj} in the $d=4$ case.

Now, using the definition of $k_2$ for the potential of a $p$-brane, eq.~\eqref{eq:brane}, one finds
\begin{equation}
    \frac{k_1}{k_2} = \frac{q(d-2)}{2-p(d-2)+d(d+k-3)}\,.
\end{equation}
As $q \leq k$ and $p< d+k-1$ it follows that 
\begin{equation}
    \frac{k_1}{k_2} < \frac{d-2}{2}\,.
\end{equation}
We see that, at least within our simplified treatment of the stringy derivation, axion kinetic energy never yields asymptotic acceleration. One cannot even achieve the borderline case $w=-(d-3)/(d-1)$ unless one starts with a higher-dimensional cosmological constant ($p= d+k-1$). One may ask whether the Kaluza-Klein scale becomes asymptotically lighter than the Hubble scale in the presence of axionic kinetic energy. We investigate this in App.~\ref{sec:axionapp}, where we also comment on an interesting heterotic example from \cite{Brinkmann:2022oxy} with the critical value $w=-1/3$ in $d=4$.

%%%%%%%%%%%%%%%%%%%%%%%%%%%%%
\subsection{Steep kination-dominated potentials}
%%%%%%%%%%%%%%%%%%%%%%%%%%%%%

We have argued that whenever an asymptotic limit is accelerating or even when the potential is somewhat steeper than necessary for acceleration, $H\geq1/R$ asymptotically. We have not concerned ourselves with what happens when the potential is very steep and the dynamics are dominated by kination. In \cite{Apers:2022cyl}, see also \cite{Rudelius:2022gbz}, it has been proposed that when the potential is too steep asymptotically, in that $V\sim \exp(-\lambda \phi)$ with $\lambda>2\sqrt{(d-1)/(d-2)}$, kination will dominate the evolution if only the scalar field drives cosmic evolution\footnote{These conclusions change in the presence of e.g.~radiation in the external dimensions \cite{Apers:2022cyl}.}, resulting in a big crunch singularity in the external directions. Combined with our results, one might conclude that only when $2\sqrt{(d+k-2)/(k(d-2))}< \lambda \leq 2\sqrt{(d-1)/(d-2)}$ does one have an asymptotic limit with $H<1/R$ and no crunch if the cosmic evolution is completely driven by a rolling scalar.

%%%%%%%%%%%%%%%%%%%%%%%%%%%%%
%%%%%%%%%%%%%%%%%%%%%%%%%%%%%
\section{Discussion}
\label{sec:discussion}
%%%%%%%%%%%%%%%%%%%%%%%%%%%%%
%%%%%%%%%%%%%%%%%%%%%%%%%%%%%

The central claim of this note is that, as per Conjecture \ref{mainconjecture}, achieving asymptotic accelerated expansion in quantum gravity is as hard as constructing a de Sitter vacuum in higher dimensions. Crucially, if one wishes to achieve asymptotic accelerated expansion in four dimensions, one should be able to construct a de Sitter vacuum in more than four dimensions. Even if this can be achieved, the KK scale falls below the cosmological curvature scale asymptotically, making the 4d description questionable.

An obvious comment in this context (not depending on Conjecture \ref{mainconjecture}) is the following: If one were certain that asymptotic acceleration is impossible in $d=3$ (by sharpening the arguments in \cite{Ooguri:2018wrx, Hebecker:2019csg} or on other ground), then long-lived metastable de Sitter vacua in $d=4$ would be ruled out. This follows by considering an $S^1$ compactification to $d=3$.

Returning to the issue of getting 4d asymptotic acceleration by compactifying dS vacua, we note that proposed constructions of de Sitter space in string theory mainly focus on compactifications to four our less spacetime dimensions. The limited work on de Sitter constructions in $d\geq 5$ (see e.g.~\cite{VanRiet:2011yc,Andriot:2022xjh,Cribiori:2023ihv}) is certainly in part due to the obvious phenomenological interest in $d=4$. However, it is also clear that it is significantly more difficult to construct higher-dimensional de Sitter vacua.

A first issue is that in five or more dimensions one cannot have SUSY with four supercharges. One must either have at least eight supercharges or completely break supersymmetry. This provides an obstacle to first stabilizing some or all of the moduli in a SUSY set-up and then achieving the de Sitter uplift as a small controlled deformation of this SUSY set-up. 

A second obstacle is that the set of available compact geometries is more limited. In particular, the ingredients for de Sitter uplifting are dimension-dependent, e.g. the Klebanov-Strassler throat necessary for the antibrane uplift in IIB requires six internal dimensions.

It then seems to us that if anything, it should be easier to construct 4D de Sitter vacua in the interior of moduli space than to achieve 4D asymptotic accelerated expansion.

In this note we have raised the point that to achieve asymptotic accelerated expansion, one requires a de Sitter vacuum in higher dimensions. If, as we suggest, the underlying reason for this is that de Sitter space and accelerated expansion are at some conceptual level `equally difficult', then one might expect that some converse statement is also true: if in the higher-dimensional theory one has (not necessarily asymptotically) some region in the scalar field space where the potential is positive and sufficiently flat, then it should be possible to construct a de Sitter vacuum in lower dimensions. In fact, \cite{Montero:2020rpl} showed that if $|V'|/V$ is sufficiently small compared to the rate at which gauge couplings change as the scalars roll then it is possible to construct a lower-dimensional de Sitter vacuum, in line with this expectation.

There exist various swampland conjectures on the profile of the scalar potential. Often an argument for a certain bound on the profile is given in the asymptotics of moduli space and it is then merely suggested that this bound may also apply to the interior of moduli space. By the results of our paper, any bound on the asymptotics has implications for the interior of moduli space in a higher-dimensional theory.

For instance, \cite{Ooguri:2018wrx, Hebecker:2018vxz} provided arguments for the refined\footnote{To be precise, the refined de Sitter conjecture states that the potential must obey either $|V'|/V\geq c$ or $V''/V \leq c'$ with $c$ and $c'$ order one coefficients. The arguments for this conjecture are presented in four spacetime dimensions, but the generalization to arbitrary dimension $d>2$ is obvious.} de Sitter conjecture $|V'|/V \geq c$ with crucial elements being the entropy of the cosmic horizon, the entropy of towers of states or simply their large number. Because of the central role of the towers of light states, these arguments clearly only apply to the asymptotics of moduli space. Unfortunately, since in this argument the value of the coefficient $c$ could not be determined, one cannot make a clear statement about de Sitter vacua in higher dimensions based on this asymptotic result alone. Later works have proposed possible asymptotic values for $c$.

In particular, the Trans-Planckian Censorship conjecture (TCC) \cite{Bedroya:2019snp,Bedroya:2019tba} proposes that $c=2/\sqrt{(d-1)(d-2)}$ asymptotically. Recently, \cite{vandeHeisteeg:2023uxj} also argued for this asymptotic behaviour based on towers of states becoming asymptotically light. By our proposal, the flattest slope one can achieve asymptotically in a string compactification exactly saturates this bound. Our results then provide an alternate argument for the asymptotic statement of the TCC. For this bound to be saturated, it is necessary that there exist long-lived metastable de Sitter vacua in the higher-dimensional theory. If there exist no higher-dimensional de Sitter vacua, then our results for the asymptotics of field space are essentially equivalent to the No Asymptotic Acceleration conjecture $c=2/\sqrt{d-2}$ of \cite{Rudelius:2021oaz,Rudelius:2021azq}. Formally, \eqref{potcodimone} (or more generally \eqref{eq:brane} for a brane of finite codimension) saturates the No Asymptotic Acceleration conjecture as $k\rightarrow +\infty$. As the number of internal directions is finite, the lower bound we obtain on $|V'|/V$ in the absence of higher-dimensional de Sitter vacua is slightly stronger than the No Asymptotic Acceleration conjecture for a given number of decompactifying dimensions.

To end, we reiterate that it would be conceptually extremely important to obtain asymptotic acceleration in string theory, even in a phenomenologically irrelevant setting. We hope that proving or disproving our AA$\Rightarrow$DS Conjecture may be a useful step in this context. A more modest goal may be to think about higher-dimensional EFTs which, without allowing for asymptotic acceleration themselves, can be compactified to models with asymptotic acceleration. If one allows for any lagrangians, this can probably be achieved. The less trivial task is to find such higher-dimensional EFTs which one may reasonably expect to arise from string theory.

%%%%%%%%%%%%%%%%%%%%%%%%%%%%%
%%%%%%%%%%%%%%%%%%%%%%%%%%%%%
\section*{Acknowledgements}
%%%%%%%%%%%%%%%%%%%%%%%%%%%%%
%%%%%%%%%%%%%%%%%%%%%%%%%%%%%
We thank Timo Weigand, Michele Cicoli, Filippo Revello, Gary Shiu, Flavio Tonioni, and Irene Valenzuela for valuable discussion. This work was supported by the Deutsche Forschungsgemeinschaft (DFG, German Research Foundation) under Germany’s Excellence Strategy EXC 2181/1 - 390900948 (the Heidelberg STRUCTURES Excellence Cluster).

%%%%%%%%%%%%%%%%%%%%%%%%%%%%%
%%%%%%%%%%%%%%%%%%%%%%%%%%%%%
\appendix
%%%%%%%%%%%%%%%%%%%%%%%%%%%%%
%%%%%%%%%%%%%%%%%%%%%%%%%%%%%

%%%%%%%%%%%%%%%%%%%%%%%%%%%%%
%%%%%%%%%%%%%%%%%%%%%%%%%%%%%
\section{Condition for asymptotic accelerated expansion}
\label{sec:expansioncond}
%%%%%%%%%%%%%%%%%%%%%%%%%%%%%
%%%%%%%%%%%%%%%%%%%%%%%%%%%%%

The purpose of this appendix is twofold. First, we  discuss asymptotic acceleration in the multifield case \cite{Calderon-Infante:2022nxb,Shiu:2023nph,Shiu:2023rxt} focusing in particular on a recent theorem due to \cite{Shiu:2023nph,Shiu:2023rxt} which shows that for certain potentials accelerated expansion is forbidden. Second, we explain in what way we will rely on this theorem in the main text. We will also argue, without proof, why we expect the theorem of \cite{Shiu:2023nph,Shiu:2023rxt} to be relevant for a broader class of potentials than those for which it has been proven. Note that for the specific case where the scalar evolution is described by gradient flow results similar to \cite{Shiu:2023nph,Shiu:2023rxt} appeared previously in \cite{Calderon-Infante:2022nxb}.

Consider a multifield scalar potential of scalars $\phi^a$. We will assume that all terms in the potential are exponential. This is expected to be true for the asymptotics of string compactifications that we are interested in. The potential is then given by
\begin{equation}
    V =  \sum_i \Lambda_i \exp(- \sum_a \gamma^i_a \phi_a)\,.
\end{equation}
When a distinction needs to be made, we will denote by $\Lambda_i^+$ ($\Lambda_i^-$) the positive (negative) $\Lambda_i$. We denote by $\gamma^{i+}_a$ ($\gamma^{i-}_a$) the $\gamma^{i}_a$ exponents of a term with a $\Lambda_i^+$ ($\Lambda_i^-$) coefficient in front.

We want to know under what conditions one asymptotically in field space achieves accelerated expansion. To do so we will rely on the results of \cite{Shiu:2023nph}, recently also extended in \cite{Shiu:2023rxt}.

Let us first focus on the case where $\Lambda_i > 0$ for all $i$. We further focus for clarity\footnote{If for some of the $\phi^a$ scalars $\gamma^{i+}_a<0$ for all terms involving those scalars, then by redefinition $\phi^a \rightarrow -\phi^a$ for those specific scalars, one recovers our situation. If a scalar has both positive and negative $\gamma_a^{i+}$, one should set $\gamma_a = 0$ for that scalar. The reason for this is intuitively clear if one considers a single scalar with potential $V(\phi)=\exp(\phi)+\exp(-\phi)$. At asymptotically late times, this scalars will stabilize at $\phi=0$ and so not contribute to the slope of the rolling of the potential at all.} on the case where all $\gamma^{i+}_a>0$. Then let $\gamma_a^+ = \min_i \gamma_a^{i+}$. It is then the case that if 
\begin{equation}\label{eq:multiaccbound}
    (\sum_a \gamma_a^+ \gamma_a^+) \geq \frac{4}{d-2}\,,
\end{equation}
there cannot be accelerated expansion at asymptotically late time \cite{Shiu:2023nph}. It immediately follows that if for any of the $\gamma_a^+$
\begin{equation}\label{eq:accbound}
    \gamma_a^+ \geq \frac{2}{\sqrt{d-2}}\equiv \gamma_\text{acc}\,,
\end{equation}
there cannot be accelerated expansion. This is the bound we will rely on for our results in Sec.~\ref{sec:branecalc}: one asks what the smallest possible $\gamma_a^+$ one can achieve in a compactification is. If even this satisfies \eqref{eq:accbound} one clearly cannot achieve accelerated expansion asymptotically.

We now need to address what happens if there are also negative terms in the potential. This involves making some technical distinctions, but the main claim we will make (without a complete proof) is that introducing additional negative terms will not help one achieve accelerated expansion. If the bound \eqref{eq:accbound} holds it will still be impossible to asymptotically achieve accelerated expansion.

As a first condition, one should have $V>0$ asymptotically despite the presence of these negative terms to even have a chance of achieving accelerated expansion. We now need to distinguish between two cases. Define $\Gamma_a^- =\max_i \gamma^{i-}_a$. 

Consider first the case when $\gamma_a^+ \geq \Gamma_a^-$ for all $a$. In this case it was shown by \cite{Shiu:2023nph} that it is still true that if \eqref{eq:multiaccbound} holds there cannot be accelerated expansion.

Finally there is the case of mixed exponents involving multiple scalars where some of the $\gamma^{i-}_a$ are larger than some of the $\gamma^{i+}_a$ and some of the $\gamma^{i-}_a$ are smaller than some of the $\gamma^{i+}_a$. In such a case the theorem of \cite{Shiu:2023nph} does not apply. While we are not able to contribute to clarifying this situation, it is our expectation that generically one of the following three possibilities will be realized in the asymptotic regime:
First, the potential can become negative if one rolls far enough, excluding asymptotic acceleration. Second, some of the moduli can become stabilized in terms of other moduli. In this case one performs the analysis of whether one has accelerated expansion in a smaller moduli space of the remaining independently rolling fields. Third, one may be able to show that in the asymptotics certain negative terms with mixed exponents can be neglected. In the second and third case, one may then hope that, after the appropriate simplification of the potential has been made, a situation arises where the theorem of \cite{Shiu:2023nph} applies. 

Let us summarize what this implies for string compactifications. 
In principle, having multiple fields can help achieving asymptotic acceleration, as for example in the case of assisted inflation \cite{Liddle:1998jc} (see also below). But we expect that this will not enable asymptotic acceleration in string compactifications. The reason is that one specific scalar field, namely the volume modulus, appears in each term of the scalar potential. We have argued that this field dependence is too steep for each term.
As a result, \eqref{eq:accbound} is satisfied for this particular field.

To elucidate what the theorem in this appendix does (and does not) imply, it may be useful to consider assisted inflation \cite{Liddle:1998jc}. Specifically, consider assisted inflation in four spacetime dimensions, for simplicity with just two scalars $\phi_1$, $\phi_2$ subject to the potential
\begin{equation}
\label{eq:asspot}
    V = \exp(-\sqrt{2}\phi_1) + \exp(-\sqrt{2}\phi_2)
\,.
\end{equation}
Each of the terms separately is too steep to permit accelerated expansion. However, consider the change of basis $\phi_1=(\phi'_1+\phi'_2)/\sqrt{2}$ and $\phi_2=(\phi'_1-\phi'_2)/\sqrt{2}$. In the new basis the potential takes the form
\begin{equation}
    V= \exp(-\phi'_1) \left[\exp(-\phi'_2)+\exp(+\phi'_2)\right]\,.
\end{equation}
Clearly, a trajectory where $\phi'_2$ is stabilized at $\phi'_2 = 0$ and only $\phi'_1$ is rolling solves the equations of motion and permits accelerated expansion. None of this however is in contradiction with the theorem. If one applies the theorem to the potential eq.~\eqref{eq:asspot} one finds $\gamma_1^+=0$ because the second term in the potential has the smallest exponent with respect to $\phi_1$, namely zero. Analogously, one also has $\gamma_2^+=0$ such that $(\sum_a \gamma_a^+ \gamma_a^+) = 0$ and the theorem does not forbid accelerated expansion. Crucially, there is no field direction in which {\it both} terms are steep, in contrast to what we expect to hold in string theory with respect to $R$.

%%%%%%%%%%%%%%%%%%%%%%%%%%%%%
%%%%%%%%%%%%%%%%%%%%%%%%%%%%%
\section{
Canonically normalized volume modulus and induced brane potential} \label{sec:appendix}
%%%%%%%%%%%%%%%%%%%%%%%%%%%%%
%%%%%%%%%%%%%%%%%%%%%%%%%%%%%

In order to calculate the slope of the potential of the $k$ dimensional volume modulus in the $d$ dimensional theory we need to canonically normalize its kinetic term.

Therefore, we consider the $(d+k)$-dimensional Einstein-Hilbert action and write the metric as
\begin{equation}
    \dd s^2 = g_{\mu\nu} \dd x^\mu \dd x^\nu + R(x)^2 \Tilde{g}_{mn} \dd y^m \dd y^n\,,
\end{equation}
where Greek indices label the external dimensions and run over $0,\dots,d-1$. Latin indices label the internal dimensions and run over $d,\dots, d+k-1$. 
Compactifying the $(d+k)$-dimensional Einstein-Hilbert term, one finds
\begin{equation} 
    S \supset M_{p,d+k}^{d+k-2} \int \dd^dx \sqrt{-g}\, R^k\left( \frac{\mathcal{R}}{2} + \frac{k(k-1)}{2} \left(\frac{\partial  R}{R}\right)^2 \right)\,,
    \label{eq:sdbd}
\end{equation}
where $M_{p,d+k}$ is the $(d+k)$-dimensional Planck mass. 

After Weyl rescaling \eqref{eq:sdbd} by $g_{\mu\nu}\to R^{-2k/(d-2)} g_{\mu\nu}$ one finds
\begin{equation}
    S \supset M_{p,d+k}^{d+k-2} \int \dd^dx \sqrt{-g} \left( \frac{\mathcal{R}}{2} - \frac{k(k+d-2)}{2(d-2)}  \left(\frac{\partial  R}{R}\right)^2  \right)\,.
\end{equation}
From here, the canonically normalized scalar $\phi$ corresponding to $R$ can easily be read off. One may also relate this scalar to the dimensionless volume of the internal space, $\mathcal{V} = (M_{p,d+k} R)^k$, as follows:
\begin{equation}
\label{eq:volumecanon}
    \mathcal{V} = \exp \left( \sqrt{\frac{k(d-2)}{k+d-2}} \,\phi \right)\,.
\end{equation}
In Sec.~\ref{sec:branecalc} we have shown that the dominant contribution to the potential of $\mathcal{V}$ is induced by a cosmological constant followed by a $(d+k-2)$-brane. 
To derive the scalar potential of $\mathcal{V}$ induced by these two effects, we calculate the contribution to the potential of a general $p$-brane. It is given by\footnote{This potential follows from dimensional analysis for effective $p$-branes in the $d+k$-dimensional theory. We have assumed that the potential does not depend on any other rolling parameters by assuming that $g_s$ is stabilized and any additional spacetime dimensions other than the $d+k$ we consider are stabilized. It is straightforward to check that allowing these to roll will not yield a flatter potential, see also the discussion in Sec. \ref{sec:limits}.}  
\begin{equation}
    \frac{V_{p-\text{brane}}}{M_{p,d}^d} = \Lambda_p \frac{M_{p,d+k}^{p+1}R^{p+1-d}}{M_{p,d}^d} = \Lambda_p \exp\left(-\underbrace{\sqrt{\frac{k(d-2)}{k+d-2}}\left( \frac{d}{d-2} -\frac{p+1-d}{k} \right)}_{\equiv \gamma}\phi\right)\,,
    \label{eq:brane}
\end{equation}
where $M_{p,d}$ is the $d$ dimensional Planck mass given by
\begin{equation}
\label{eq:planckconversion}
    M_{p,d}^{d-2} = M_{p,d+k}^{d+k-2} R^k\,,
\end{equation}
and $\Lambda_p$ a positive constant.

%%%%%%%%%%%%%%%%%%%%%%%%%%%%%
%%%%%%%%%%%%%%%%%%%%%%%%%%%%%
\section{Scale-separation analysis with axionic kinetic energy}\label{sec:axionapp}
%%%%%%%%%%%%%%%%%%%%%%%%%%%%%
%%%%%%%%%%%%%%%%%%%%%%%%%%%%%

In Sec.~\ref{sec:externaleffects} we discussed the asymptotic behaviour in the decompactification limit in the presence of a rolling axion. Similarly to the analysis in Sec.~\ref{sec:EFTbreakdown}, one may ask also in this case whether scale separation is lost, i.e.~whether~$H\geq1/R$.

We start by thinking in $d$-dimensional effective theory, suppressing powers of $M_{p,d}$ for simplicity. Using \eqref{eq:axionkin} and \eqref{eq:volumecanon}, the (dualized) scalar kinetic energy may be rewritten according to
\begin{equation}
    \exp(2k_1\phi){\cal H}_{d-1}^2 \sim \left(M_{p,d+k}\,R\right)^{2q}
    \frac{1}{a^{2(d-1)}}\,.
    \label{hrr}
\end{equation}
Now recall that, when the axion drives acceleration, kinetic and potential energy decay at the same rate. The Friedmann equation then implies that \eqref{hrr} scales like $H^2$. Moreover, it is well known that an equation-of-state parameter $w$ implies $H^2\sim 1/a^n$ with $n=(d-1)(1+w)$. Thus, \eqref{hrr} scales like $1/a^n$ and it follows that $a \sim (M_{p,d+k}\,R)^\frac{2q}{2(d-1)-n}$. 

Now we are ready to compare the $d$-dimensional parameter $H$ to the higher-dimensional parameter $1/R$. At this point it is necessary to restore $M_{p,d}$ and use \eqref{eq:planckconversion}. The ratio of Hubble and KK scale then takes the form 
\begin{equation}
    \frac{H}{1/R} \sim \frac{M_{p,d}/a^{n/2}}{1/R} = (M_{p,d+k}\,R)^{\frac{d+k-2}{d-2}}\,a^{-n/2} \sim (M_{p,d+k}\,R)^{\frac{d+k-2}{d-2}-\frac{qn}{2(d-1)-n}}\,.
\end{equation}
Hence, scale separation is lost if 
\begin{equation}
    (d+k-2)(2(d-1)-n) \geq nq(d-2) \,.
    \label{eq:axionss}
\end{equation}
For the borderline case of $n=2$ (or equivalently $w=-(d-3)/(d-1)$) one finds that scale separation is lost for
\begin{equation}
   d+k-2   \geq q \,,
\end{equation}
which is always true in $d>2$ since $q\leq k$. This then agrees with the analysis in Sect.~\ref{sec:EFTbreakdown}.

Interestingly, \cite{Brinkmann:2022oxy} also provides an example in $d=4$ where the attractor has $w=-1/3$ based on the heterotic compactifications of \cite{Cicoli:2013rwa}. Crucially, this set-up involves a rolling axiodilaton such that our preceding analysis of when $H>1/R$ is not applicable to this case. It would be interesting to analyze this case further and see whether $H>1/R$ asymptotically. Let us stress however that since $w=-1/3$ asymptotically, this case does not accelerate asymptotically per our definition and this example is not a counterexample to our conjecture even if it does not stem from a higher-dimensional de Sitter vacuum.

%%%%%%%%%%%%%%%%%%%%%%%%%%%%%
%%%%%%%%%%%%%%%%%%%%%%%%%%%%%

\bibliographystyle{JHEP}
\bibliography{refs}

%%%%%%%%%%%%%%%%%%%%%%%%%%%%%
%%%%%%%%%%%%%%%%%%%%%%%%%%%%%
\end{document}